\documentclass[sigconf]{aamas}

\usepackage{balance} 
\usepackage{soul}
\usepackage{xcolor}
\usepackage{tikz}
\usepackage[font=small]{caption}
\usepackage{subcaption}

\setcopyright{ifaamas}
\acmConference[AAMAS '22]{Proc.\@ of the 21st International Conference
on Autonomous Agents and Multiagent Systems (AAMAS 2022)}{May 9--13, 2022}
{Auckland, New Zealand}{P.~Faliszewski, V.~Mascardi, C.~Pelachaud,
M.E.~Taylor (eds.)}
\copyrightyear{2022}
\acmYear{2022}
\acmDOI{}
\acmPrice{}
\acmISBN{}

\title[]{Learning Generalizable Multi-Lane Mixed-Autonomy Behaviors in Single Lane Representations of Traffic}

\author{Abdul Rahman Kreidieh}
\affiliation{
  \institution{University of California, Berkeley}
  \city{Berkeley}
  \state{California}
  \country{United States}}
\email{aboudy@berkeley.edu}

\author{Yibo Zhao}
\affiliation{
  \institution{University of California, Berkeley}
  \city{Berkeley}
  \state{California}
  \country{United States}}
\email{brentzhao@berkeley.edu}

\author{Samyak Parajuli}
\affiliation{
  \institution{University of California, Berkeley}
  \city{Berkeley}
  \state{California}
  \country{United States}}
\email{samyak.parajuli@berkeley.edu}

\author{Alexandre Bayen}
\affiliation{
  \institution{University of California, Berkeley}
  \city{Berkeley}
  \state{California}
  \country{United States}}
\email{bayen@berkeley.edu}

\begin{abstract}
Reinforcement learning techniques can provide substantial insights into the desired behaviors of future autonomous driving systems. By optimizing for societal metrics of traffic such as increased throughput and reduced energy consumption, such methods can derive maneuvers that, if adopted by even a small portion of vehicles, may significantly improve the state of traffic for all vehicles involved. These methods, however, are hindered in practice by the difficulty of designing 
efficient and accurate models of traffic, as well as the challenges associated with optimizing for the behaviors of dozens of interacting agents. In response to these challenges, this paper tackles the problem of learning generalizable traffic control strategies in simple 
representations of 
vehicle 
driving dynamics. 
In particular, we look to mixed-autonomy ring roads as 
depictions of instabilities 
that result in the formation of congestion. Within this problem, we design a curriculum learning paradigm that exploits the natural extendability of the network 
to effectively learn behaviors that reduce congestion over long horizons. Next, we study the implications of 
modeling 
lane changing on the transferability of policies. Our findings suggest that introducing lane change behaviors that even approximately match trends in more complex systems can significantly improve the generalizability of subsequent learned models to more accurate multi-lane models of traffic.
\end{abstract}

\keywords{Reinforcement Learning, Social Simulation, Traffic Control}

\newcommand{\BibTeX}{\rm B\kern-.05em{\sc i\kern-.025em b}\kern-.08em\TeX}

\begin{document}


\pagestyle{fancy}
\fancyhead{}


\maketitle 


\section{Introduction}

Reinforcement learning (RL) methods will likely play a significant role in the design of future autonomous driving strategies. As is the case in several cooperative/competitive settings~\cite{mordatch2018emergence, leibo2017multi, bansal2017emergent, silver2016mastering}, these methods may provide significant insights into the nature of desirable interactions by \emph{automated vehicles} (AVs). In particular, if properly applied, these methods may enhance our understanding of the quality of policies needed for socially optimal and energy-efficient driving to emerge. As such, efficient learning methods are needed to ensure that such systems succeed.

Due to the complexity of deploying RL methods in real-world, safety-critical environments, studies have focused primarily on analyzing such methods in simulation. In~\cite{wu2021flow}, for instance, the authors demonstrate the reinforcement learning policies can match and even exceed the performance of controllers already demonstrated to provide benefits in real-world settings~\cite{stern2018dissipation}. This has been done for more complex problems as well, with other studies showing performative benefits of decentralized autonomous vehicles trained to operate in lane-restricted bottlenecks~\cite{vinitsky2020optimizing} and intersections~\cite{wu2019dcl}, among others~\cite{wu2017emergent,vinitsky2018benchmarks,cui2021scalable}. However, these methods are often hindered by the process of modeling and simulating large-scale transportation networks and the difficulty of jointly optimizing for the behavior of multiple vehicles within these settings. Thus, robust learning methods in more simplified traffic control problems are needed to address this.

In this paper, we demonstrate that policies learned in simplified and computationally efficient ring road settings, if properly defined, can effectively learn behaviors that generalize to more complex problems. To enable proper generalization, we identify two limiting factors that exacerbate the dynamical mismatch between the two problems and describe methods for addressing each of them while maintaining the relative simplicity and efficiency of the original task. For one, to address mismatches that arise from variations in the boundary conditions, we construct a curriculum learning paradigm that scales the performance of policies learned to larger rings, where boundaries pose less of a concern. Next, to introduce perturbations that arise for lane changes and cut-ins, we introduce a simple approach for simulating such disturbances in single-lane settings and study the implications of the aggressive and accuracy of such disturbances on the resultant policy.

\begin{figure*}
    \includegraphics[width=\textwidth]{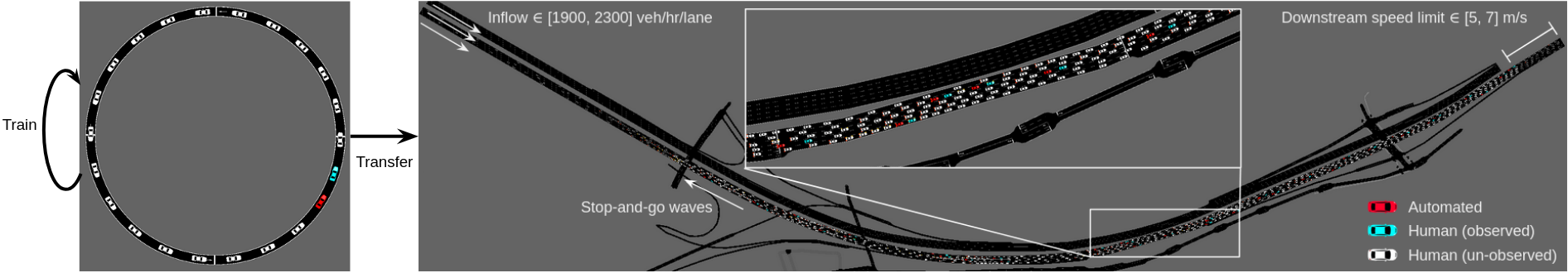}
    \caption{An illustration of the explored problem and transfer learning procedure. This paper aims to learn policies in simplified ring road settings transferable to more complex, real-world tasks. The target network of interest, seen in the right, is a simulated section of the I-210 network in Los Angeles, CA, in which restrictions to the outflow conditions result in the formation of stop-and-go congestion.}
    \label{fig:transfer-setup}
\end{figure*}

We validate the performance of our approach on a calibrated model of the I-210 network in Los Angeles, California (Figure~\ref{fig:transfer-setup}). Our findings suggest that learning in larger rings via curricula and introducing simple, random lane change events greatly generalizes the resultant congestion-smoothing behaviors. An exploration of the rate with which lane change occurs also provides insights into the types of lane changes needed to generalize to other similar settings.

The primary contributions of this paper are:

\begin{itemize}
    \item We design a curriculum learning paradigm consisting of rings of increasing length and number of AVs to efficiently learn multiagent vehicle interactions in large-scale settings.
    \item We explore methods for modeling lane changes in single-lane problems that improve the generalizability of learned policies to more realistic multi-lane settings.
    \item We demonstrate the relevance of the above to methods in learning effective congestion-mitigation policies in large-scale, flow-restricted networks.
\end{itemize}

The remainder of this paper is organized as follows. Section~\ref{sec:prelims} introduces relevant concepts and terminology and their relation to the present paper. Section~\ref{sec:method} discusses the problem statement and the techniques utilized to improve the generalizability of policies learned in ring roads. Section~\ref{sec:results} presents numerical results of the proposed methods and provides insights into the degree of generalizability achieved. Finally, Section~\ref{sec:conclusion} provides concluding remarks and discusses potential avenues for future work.



\section{Related Work} \label{sec:prelims}

\subsection{Addressing Congestion in Ring Roads}

Ring roads model the behavior of $n$ vehicles in a circular track of length $L$ with periodic boundary conditions. Within these models, the evolution of the state (or position $x_i, i \in {1,\dots, n}$) of each vehicle is represented via the system of ordinary differential equations:
\begin{equation}
	\begin{cases}
	\ddot{x}_i(t) = f\left(v_i(t), v_{i+1}(t), h_i(t) \right) & i = 1, \dots, n-1\\
	\ddot{x}_n(t) = f\left(v_n(t), v_1(t), h_n(t) \right) \\
	x_i(t_0) = x_0^i & i = 1, \dots, n \\
	\dot{x}_i(t_0) = v_0^i & i = 1, \dots, n
	\end{cases}
	\label{eq:ring}
\end{equation}
where $f(\cdot)$ is a car-following model that mimics the acceleration of human driver $i$ as a function of their ego speed $v_i(t)$, the speed of their leader, and the bumper-to-bumper headway between the two $h_i(t)$, and $x_0^i$ and $v_0^i$ are the initial position and speed of the vehicle, respectively. \st{TODO: example.}

Single-lane ring roads have served as a proxy for studying traffic instabilities for several decades. Within these problems, instabilities appear as a result of behaviors inherent to human driving. In particular, due to \emph{string instabilities}~\cite{swaroop1996string} in human driving behaviors, small fluctuations in driving speeds produce higher amplitude oscillations by following vehicles as they break harder to avoid unsafe settings. This response perturbs the system from its desirable \emph{uniform flow} equilibrium, in which all vehicles are equidistant and drive at a constant speed, and results in the formation of stop-and-go waves common to highway networks, thereby contributing to the reduced throughput and efficiency of the network. This effect has been empirically validated by the seminal work of Sugiyama et al.~\cite{sugiyama2008traffic}, and in simulation may be reconstructed through the use of popular car-following models (e.g~\cite{treiber2000congested,bando1995dynamical}).

The presence of string instabilities in human drivers highlights a potential benefit for upcoming autonomous driving systems. In particular, by responding to string instabilities in human driving, \emph{automated vehicles} (AVs) at even low penetrations can reestablish uniform flow driving, improving throughput and energy-efficiency in the process. Several studies have attempted to establish appropriate policies~\cite{sun2018stability, cui2017stabilizing, wang2019chain}. Most relevant to the present paper, studies have also looked into the characteristics of emergent phenomena learned via deep reinforcement learning in mixed-autonomy ring roads with socially optimal objectives~\cite{wu2017emergent}. The behaviors, however, lend little insight into the nature of desired behaviors needed to achieve socially optimal driving in real, large-scale networks. We highlight the limiting factors in Section~\ref{sec:limiting} and provide methods for addressing them in Section~\ref{sec:curriculum}~and~\ref{sec:modeling}. We then discuss the implications of these factors on learned behaviors in Section~\ref{sec:transfer-performance}.

\subsection{Reinforcement Learning and MDPs}

RL problems are studied as a \textit{Markov decision problem} (MDP)~\cite{bellman1957markovian}, defined by the tuple: $(\mathcal{S}, \mathcal{A}, \mathcal{P}, r, \rho_0, \gamma, T)$, where $\mathcal{S} \subseteq \mathbb{R}^n$ is an $n$-dimensional state space, $\mathcal{A} \subseteq \mathbb{R}^m$ an $m$-dimensional action space, $\mathcal{P} : \mathcal{S} \times \mathcal{A} \times \mathcal{S} \to \mathbb{R}_+$ a transition probability function, $r : \mathcal{S} \to \mathbb{R}$ a bounded reward function, $\rho_0 : \mathcal{S} \to \mathbb{R}_+$ an initial state distribution, $\gamma \in (0,1]$ a discount factor, and $T$ a time horizon. For multiagent problems in particular, these problems are further expressed as \emph{Markov games}~\cite{littman1994markov}, consisting a collection of state $\mathcal{S}_1, \dots, \mathcal{S}_k$ and action sets $\mathcal{A}_1, \dots, \mathcal{A}_k$ for each agent in the environment.

In a Markov game, an \textit{agent} $i$ is in a state $s_i \in \mathcal{S}_i$ in the environment and interacts with this environment by performing actions $a_i \in \mathcal{A}_i$. The agent's actions are defined by a policy $\pi_{\theta} : \mathcal{S}_i \times \mathcal{A}_i \to \mathbb{R}_+$ parametrized by $\theta$. The objective of the agent is to learn an optimal policy: $\theta^* := \text{argmax}_{\theta} \eta(\pi_{\theta})$, where $\eta(\pi_{\theta}) = \sum_{t=0}^T \gamma^t r_t$ is the expected discounted return. In the present article, these parameters are updated using direct policy gradient methods~\cite{williams1992simple}, and in particular the TRPO algorithm~\cite{schulman2015trust}.

\subsection{Transfer Learning}

Transfer learning techniques in reinforcement learning provide methods of leveraging experiences acquired from training in one task to improve training on another~\cite{taylor2009transfer}. These tasks may differ from the perspective of the agent (e.g., the observation the agent perceives or the actions it may perform) or other components of the MDP structure (e.g., the transition probability). Standard transfer learning practices include sharing policy parameters and state-action pairs between tasks. For a survey of transfer learning techniques, we refer the reader to~\cite{taylor2009transfer,pan2009survey}.

The notion of transferring knowledge from policies learned on ring roads to more complex tasks has been explored in the past. In particular, the work of~\cite{kreidieh2018dissipating} explores the transferability of policies learned on a closed ring road network to open highway networks with instabilities arising from an on-ramp merge. It finds that policies trained on ring road networks exhibiting similar network densities as their target network performs well in single-lane merges without exposure to the network. This work, however, provides limited success in generating meaningful control strategies for multi-lane highway settings, where additional approximations for the dynamics of lane-changing are needed. We introduce methods for addressing this challenge in the following section.



\section{Experimental Setup} \label{sec:method}

In this section, we introduce the explored mixed-autonomy control problem and the features within this problem that limits its generalizability to complex tasks. We then introduce methods to improve the scalability and transferability of learned policies.

\subsection{Problem Definition} \label{sec:problem-definition}

We begin by establishing the training environment under which policies learn to regulate congestion in simplified ring road representations of traffic. This is an extension of the specifications provided in~\cite{wu2021flow}, with modifications aimed at improving the performance of learning methods in multiagent settings.\\[-5pt]

\noindent \textbf{Network/actions:} We define the mixed-autonomy ring road problem as an extension of the model depicted in Eq.~\eqref{eq:ring}, in which the actions of decentralized agents dictate the desired accelerations by AVs. Let $\mathcal{S}_{AV} \subset \{1, \dots, n\}$ consist of the set of AVs whose actions are dictated by a learning agent, or policy, $\pi_{\theta}$. The updated system of ordinary differential equations dictating the dynamics of the network is then:
\begin{equation}
	\begin{cases}
	\ddot{x}_i(t) = f\left(v_i(t), v_{i+1}(t), h_i(t) \right) & i = 1, \dots, n-1\ |\ i \not\in \mathcal{S}_{AV} \\
	\ddot{x}_i(t) = \pi_{\theta} \left(s_i(t)\right) & i \in \mathcal{S}_{AV} \\
	\ddot{x}_n(t) = f\left(v_n(t), v_1(t), h_n(t) \right)
	\end{cases}
	\label{eq:ring-av}
\end{equation}
where $s_i(t)$ is the state of AV $i$ at time $t$, defined later in this section. For the car-following model $f$, we consider the \emph{Intelligent Driver Model} (IDM)~\cite{treiber2000congested}, a popular model for accurately simulating string-instabilities in human driving. Through this model, the acceleration for a  vehicle $\alpha$ is defined by its headway $h_\alpha$, velocity $v_\alpha$, and relative velocity with the preceding vehicle $\Delta v_\alpha = v_l - v_\alpha$ as:
\begin{equation}
f(v_\alpha, v_l, h_\alpha) = a \left[ 1 - \left( \frac{v_\alpha}{v_0} \right)^\delta - \left( \frac{s^*(v_\alpha, \Delta v_\alpha)}{h_\alpha} \right)^2 \right] + \epsilon
\end{equation}
where $\epsilon$ is an exogenous noise term designed to mimic stochasticity in driving, and $s^*$ is the desired headway of the vehicle denoted by:
\begin{equation}
s^*(v_\alpha, \Delta v_\alpha) = s_0 + \text{max}\left(0, v_\alpha T + \frac{v_\alpha \Delta v_\alpha}{2 \sqrt{ab}} \right)
\end{equation}
and $h_0$, $v_0$, $T$, $\delta$, $a$, $b$ are given parameters provided in Table~\ref{table:model-params}.\\[-5pt]

\noindent \textbf{Observations:} The observation space of the learning agent consists of locally observable network features. This includes the speed $v_{i,\text{lead}}$ and bumper-to-bumper headway $h_i$ of the vehicle immediately preceding the automated vehicle $i$ and its ego speed $v_i$, extended over a given number of frames $N$ and sampling rate $\Delta t$. The final state $s_i(t)$ at time $t$ is:
\begin{equation}
    \resizebox{.9\hsize}{!}{$
        \begin{aligned}
        s_i(t) = (&v_\alpha(t), h_\alpha (t), v_l(t),\\
        &v_\alpha(t-\Delta t), h_\alpha (t \Delta t), v_l(t\Delta t),\\
        &\vdots\\
        &v_\alpha(t-(N-1)\Delta t), h_\alpha (t -(N-1)\Delta t), v_l(t-(N-1)\Delta t))        
        \end{aligned}
    $}
\end{equation}
where $N=5$ and $\Delta t=2$.\\[-5pt]

\noindent \textbf{Rewards:} We choose a reward function that aims to direct the flow of traffic towards its uniform flow speed while regularizing the actions performed by the AVs. Moreover, due to the difficulty of credit assignment when multiple 
agents interact with one another, we choose to represent this behavior as a function of features local to the individual AVs. The reward $r_i(t)$ of agent $i$ at time $t$ is:
\begin{equation}
    r_i(t) = -c_1 \left(\dot{x}_i(t) - V_\text{eq}(\rho)\right)^2 - c2 \left(\ddot{x}_i(t)\right)^2
\end{equation}
where $V_\text{eq}(\rho)$ is the uniform flow speed at a given network density $\rho$ (see~\cite{wu2017emergent} for more details), and the constants are set to $c_1 = 0.005$ and $c_2 = 0.1$.

\begin{table}[]
\centering
\begin{footnotesize}
\begin{tabular}{|l|c|c|c|c|c|c|c|}
\hline
& \multicolumn{7}{c|}{Intelligent Driver Model (IDM)} \\ \hline
\textbf{Parameter} & $v_0$ & $T$ & $a$ & $b$ & $\delta$ & $s_0$ & $\epsilon$ \\ \hline
\textbf{Value} & $30$ & $1$ & $1.3$ & $2.0$ & $4$ & $2$ & $\mathcal{N}(0,0.3)$ \\ \hline
\end{tabular}
\end{footnotesize}
\caption{Model parameters, set to match~\cite{lee2021integrated}}
\label{table:model-params}
\end{table}

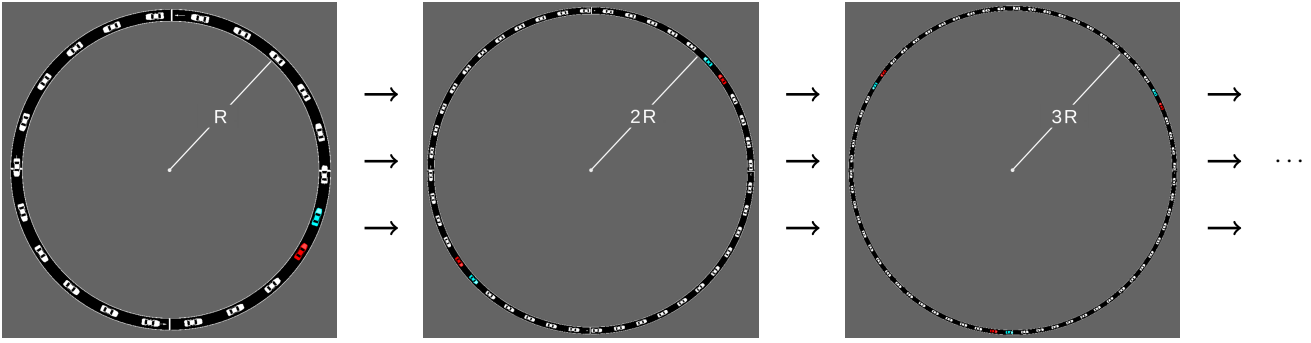
\begin{figure*}
    \begin{subfigure}[b]{0.25\textwidth}
        \includegraphics[width=\textwidth,height=\textwidth]{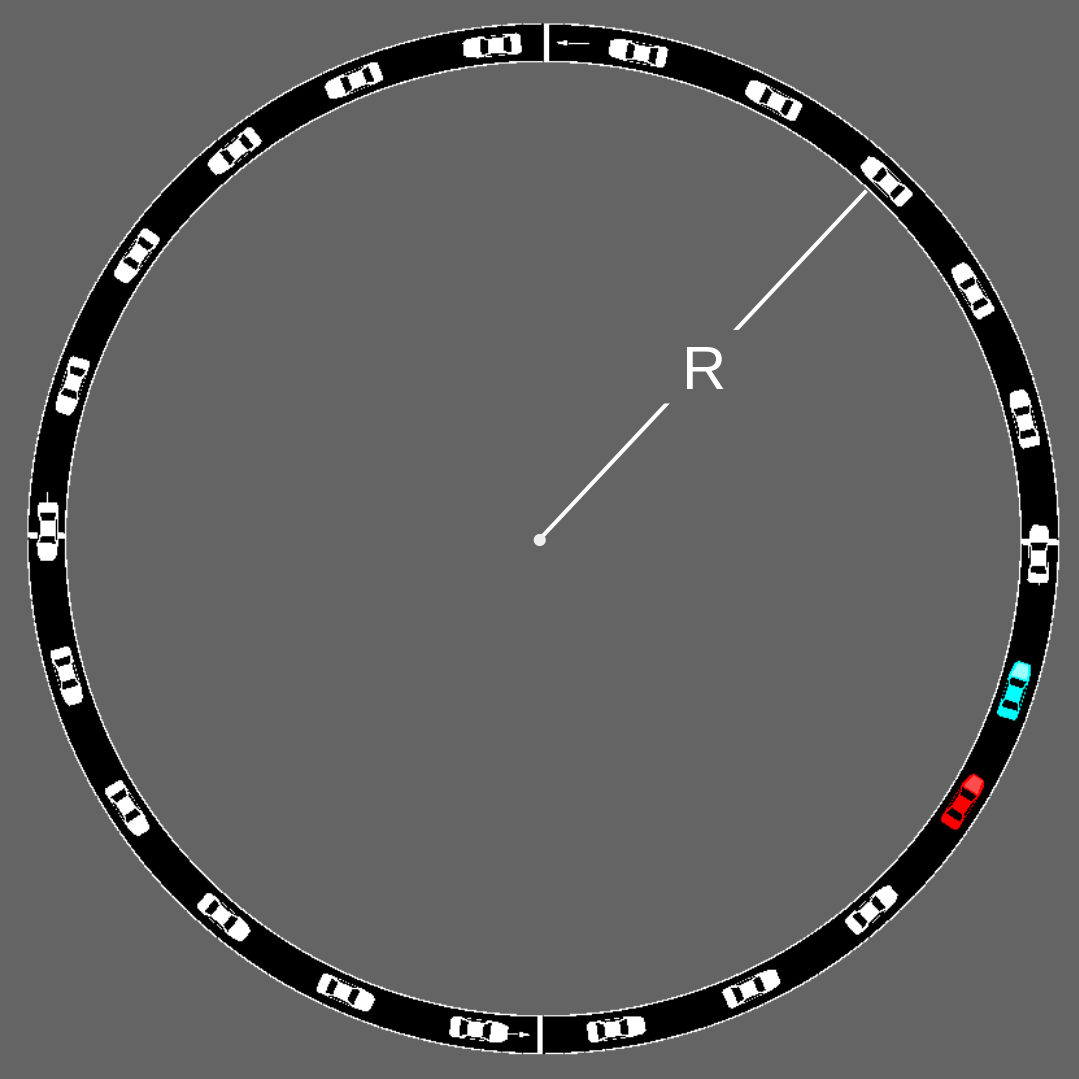}
    \end{subfigure}
    \ \
    \begin{subfigure}[b]{0.025\textwidth}
        \begin{tikzpicture}
            \node[] () at (0, 0) {\footnotesize \textcolor{white}{.}};
            \draw[->, line width=1] (0, 3\linewidth) -- (\linewidth, 3\linewidth);
            \draw[->, line width=1] (0, 5\linewidth) -- (\linewidth, 5\linewidth);
            \draw[->, line width=1] (0, 7\linewidth) -- (\linewidth, 7\linewidth);
        \end{tikzpicture}
    \end{subfigure}
    \ \ \ \ \
    \begin{subfigure}[b]{0.25\textwidth}
        \includegraphics[width=\textwidth,height=\textwidth]{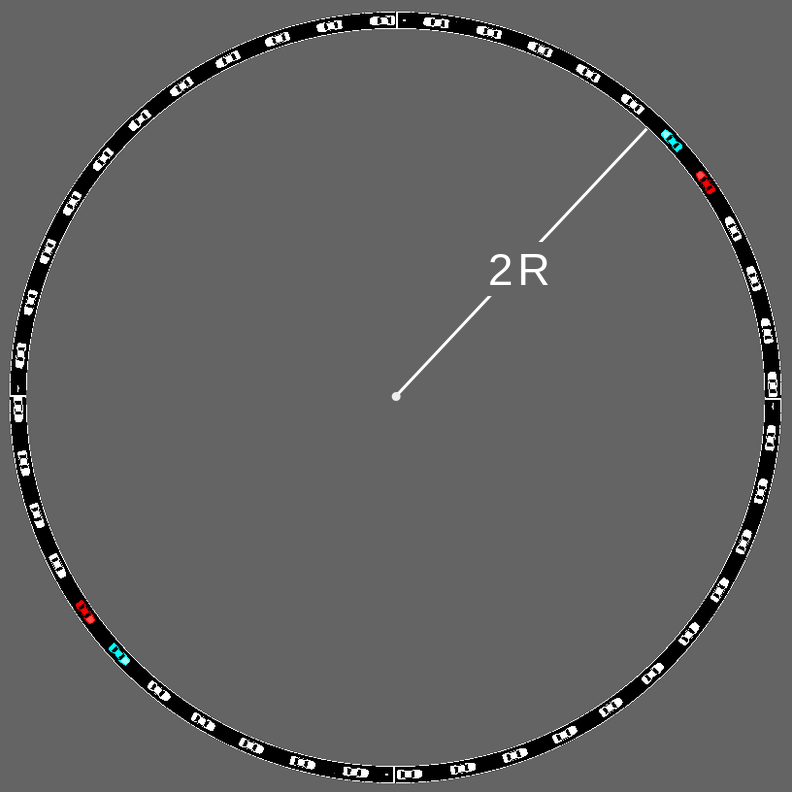}
    \end{subfigure}
    \ \
    \begin{subfigure}[b]{0.025\textwidth}
        \begin{tikzpicture}
            \node[] () at (0, 0) {\footnotesize \textcolor{white}{.}};
            \draw[->, line width=1] (0, 3\linewidth) -- (\linewidth, 3\linewidth);
            \draw[->, line width=1] (0, 5\linewidth) -- (\linewidth, 5\linewidth);
            \draw[->, line width=1] (0, 7\linewidth) -- (\linewidth, 7\linewidth);
        \end{tikzpicture}
    \end{subfigure}
    \ \ \ \ \
    \begin{subfigure}[b]{0.25\textwidth}
        \includegraphics[width=\textwidth,height=\textwidth]{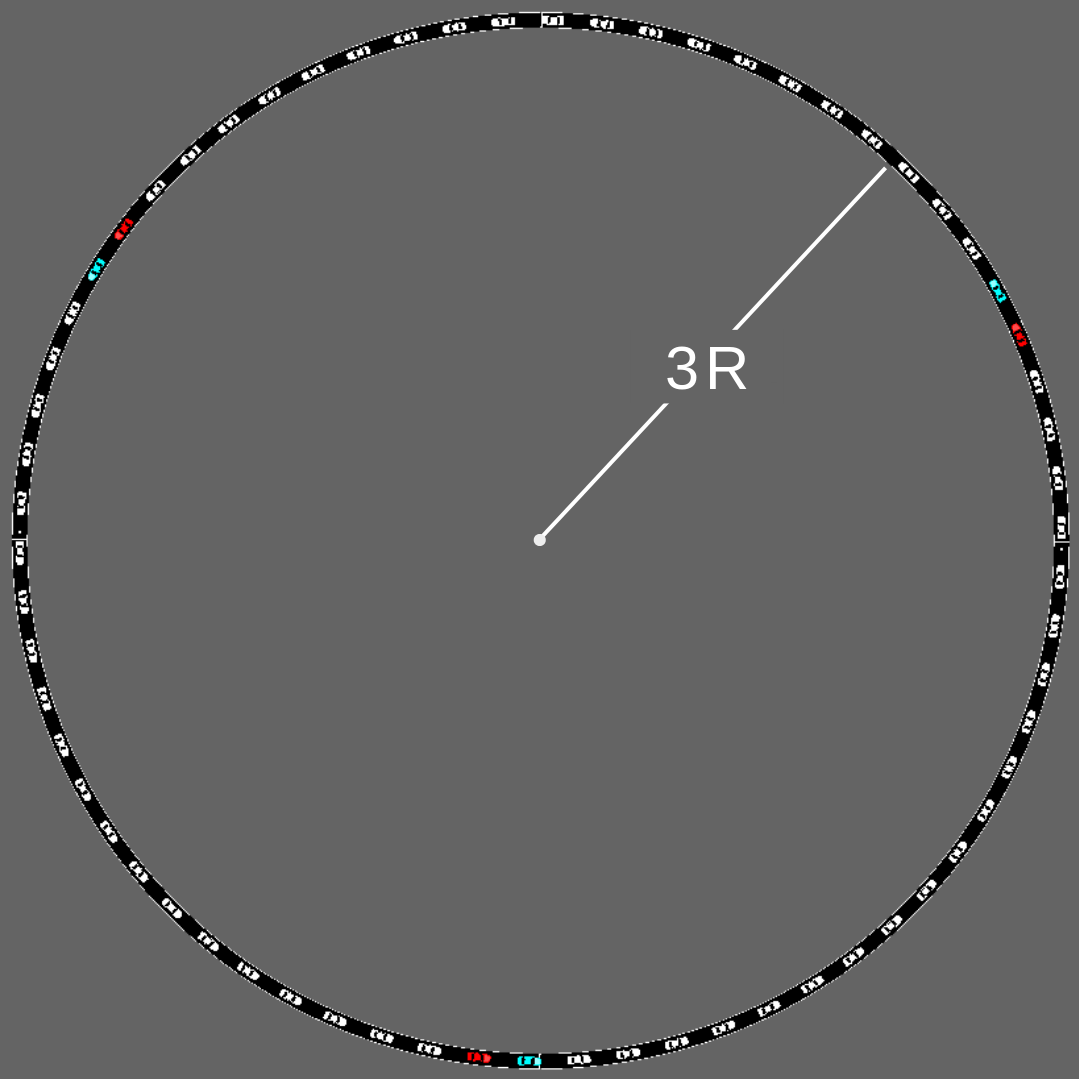}
    \end{subfigure}
    \ \
    \begin{subfigure}[b]{0.025\textwidth}
        \begin{tikzpicture}
            \node[] () at (0, 0) {\footnotesize \textcolor{white}{.}};
            \draw[->, line width=1] (0, 3\linewidth) -- (\linewidth, 3\linewidth);
            \draw[->, line width=1] (0, 5\linewidth) -- (\linewidth, 5\linewidth);
            \draw[->, line width=1] (0, 7\linewidth) -- (\linewidth, 7\linewidth);
        \end{tikzpicture}
    \end{subfigure}
    \ \ \ \ \
    \begin{subfigure}[b]{0.05\textwidth}
        \begin{tikzpicture}
            \node[] () at (0, 0) {\footnotesize \textcolor{white}{.}};
            \node[] () at (0, 2.5\linewidth) {\large $\dots$};
        \end{tikzpicture}
    \end{subfigure}
    \caption{An illustration of the curriculum learning procedure. We attempt to learn initial policies within a small ring with a single autonomous vehicles, and consecutively transition the policy to problems with increased length and number of AVs to improve the structure of the solution to relaxed boundary conditions and interacting between AVs. The length of the rings in each problem are chosen to approximately match the expected range densities within the target multi-lane problem.}
    \label{fig:curriculum-setup}
\end{figure*}

\subsection{Limitations to Generalizability} \label{sec:limiting}

The process of learning mixed-autonomy traffic regulation policies within simulated ring roads presents several notable benefits in the context of RL. For one, the simplicity of the proposed dynamics renders the problem easy to reconstruct and computationally efficient to simulate over discrete time segments. This is particularly beneficial within the field of RL, which relies largely on reproducibility and benchmarking to progress, and within which existing methods often require millions of interactions to generate meaningful behaviors. In addition, notions of stability and social optimally render the definition of the reward, or objective function, relatively straightforward. This is in contrast to RL studies in more complex mixed-autonomy settings, which rely primarily on heuristic insights for engineering reward functions that handle, for example, the presence on and off-ramps, insights that may not be transferable to other domains. The question, nevertheless, remains: Are policies learned within these settings meaningful/generalizable to complex, multifeatural networks?

To answer this question, we identify two features limiting ring road policies' applicability to multi-lane, open highway networks. We address these limitations in the following subsections.

\subsubsection{Boundary conditions}
The first of these challenges considers the effect of periodic boundary conditions unique to closed (circular) networks. In smaller ring networks, similar to those previously studied, this boundary strongly couples the actions performed by individual AVs and the long-term responses of all other vehicles in the network, including those in front of the AV. In real-world networks, however, the effect vehicles have on traffic is unidirectional, propagating against the flow of traffic. The presence of this condition, as such, introduces potential optimal behaviors that limit the potential transferability of any learned behavior.

\subsubsection{Perturbations by adjacent lanes} In addition to the effects of boundary conditions, perturbations to the observed state induced by lane changes are also absent in the single-lane ring. These perturbations include sudden reductions in headway resulting from overtaking actions and temporal fluctuations in vehicle densities and destabilize the learned policy if not adequately captured by the source task. As such, solutions need to alleviate the shift between the two domains.

\subsection{Learning Scalable Behaviors via Curricula} \label{sec:curriculum}

We begin by addressing the challenges of learning transferable open-network behaviors in environments with closed-loop / periodic boundary conditions. As mentioned in the previous subsection, the presence of this boundary condition produces strong couplings between actions by an AV and the behaviors of vehicles directly in front of it. This effect is in direct opposition to the natural convective stability of open networks, which likely results in the formation of undesirable emergent behaviors.

Larger diameter ring roads with additional automated vehicles, as noted in~\cite{orosz2009exciting}, can help dilute these boundary effects. Learning in these more complex settings, however, poses a significant challenge to existing multiagent RL algorithms. In particular, the increasingly delayed effects that individual actions have on the global efficiency of these networks, coupled with the effects of further non-stationary and credit assignment problems, hinder the progression of gradient-based methods as they attempt to traverse the space of possible solutions. As a result, features within the learning procedure as simple as policy initializations heavily dictate the nature of the converged behavior in these larger settings.

To improve the performance of policies in larger ring problems, we design a curriculum learning paradigm that exploits the extendability of the problem and the similarity in solutions between rings of similar magnitudes. This approach is depicted in Figure~\ref{fig:curriculum-setup}. Within this paradigm, policies to be trained in mixed-autonomy rings of length $nL$ with $n$ AVs are initially pretrained in a ring of length $L$ with $1$ AV for a total of $n_\text{pretrain}$ epochs. This initial problem is relatively simple to solve but does not sufficiently generalize the $n$-AV problem due to the vast difference in effects from the periodic boundary. However, the solution serves as a beneficial initialization for scaling the performance of policies learned in the $2$-AV settings, which share closer boundary effects. As such, a new pretrain policy is learned on the $2$-AV ring using the policy learned prior as a warm start, and this process is repeated for $n-1$ iterations until a proper policy initialization may be provided to the $n$-AV problem, which is then trained for a total of $n_\text{train}$ epochs.

\subsection{Modeling Disturbances by Lane Changes} \label{sec:modeling}

\begin{figure}
    \begin{subfigure}[b]{\linewidth}
        \centering
        \includegraphics[width=\linewidth]{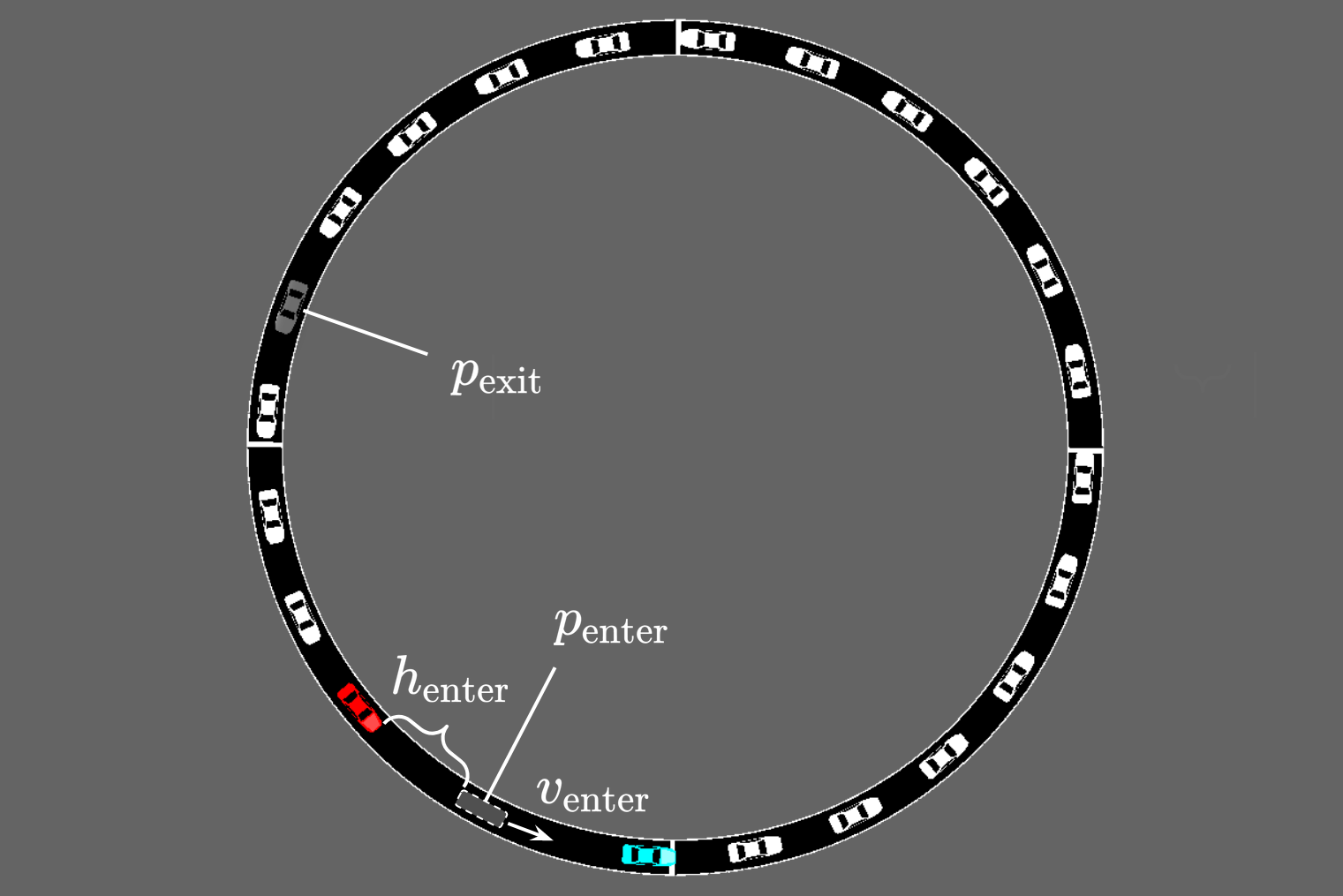}
    \end{subfigure}
    \caption{We simulate disturbances arising from lane changes in the single lane setting by stochastically placing and removing vehicles within the ring. These events are characterized by their occurrence probability as well as the initial conditions for the entry vehicles.}
    \label{fig:lane-change-setup}
\end{figure}

In this subsection, we attempt to address the second of the challenges presented in Section~\ref{sec:limiting}, namely the shift in domains induced by the presence of lane changes in real-world tasks. To resolve this concern,
we attempt to model the effects of lane changes from the perspective of individual lanes. In particular, we take inspiration from the work of~\cite{wu2017multi} and model lane changes as stochastic insertions and deletions of vehicles within the ring. As seen in Figure~\ref{fig:lane-change-setup}, the insertions, which model lane-change-in actions, are characterized by their event probability $p_\text{enter}$ and the entry speed $v_\text{enter}$ and headway $h_\text{enter}$ with respect to the trailing vehicle, and deletions, which model lane-change out events, are characterized by their event probability $p_\text{exit}$.

In modeling the insertion/deletion event probabilities and entry conditions, we wish to design simple representations that sufficiently account for the effects of perturbations to ensure effective generalizability. Random lane change events represent the simplest approach for modeling such effects on automated vehicles. For vehicles whose headway permits a safe lane change, the lane change events are modeled as identical independent random variables of a Bernoulli distribution. Through such an approach, the components of the lane change events at a given time are defined as:

\begin{equation}
    p_\text{exit} \sim \frac{\mathbb{E}_{out}}{n_{t}}
    \label{eq:pexit}
\end{equation}
\begin{equation}
    p_\text{enter} \sim \frac{\mathbb{E}_{in}}{n_{t}}
    \label{eq:penter}
\end{equation}
\begin{equation}
    v_\text{enter} \sim \text{Average speed across all vehicles.}
\end{equation}
\begin{equation}
    h_\text{enter} \sim \text{headway} / 2
\end{equation}
where $\mathbb{E}_{out}$ and $\mathbb{E}_{in}$ are the hyperparameters for the expected number of lane change events at each time step. $n_{t}$ is the number of vehicles who can safely experience an overtaking event at time $t$.

The use of random lane change actions, in addition to its design simplicity, absolves the modeling procedure of its dependence on data and estimation. As a result, policies learned through such a lane change model are not biased by small datasets or poor performing estimations and instead must learn to perform generally well under any potential lane change event, aggressive or otherwise. In the context of transfer learning, this approach can be considered closest to domain randomization~\cite{tobin2017domain}. 
We explore the effects of the $\mathbb{E}_{out}$ and $\mathbb{E}_{in}$ terms in the following section.

\subsection{Simulation and Training}

Simulations of the ring-road problem are performed via a discretized variant of the model depicted in Eq.~\ref{eq:ring-av} with a simulation time step of $0.2$ sec/step. We consider four variants of the ring environment, which we characterized by the number of controlled agents $N_{AV} \in \{1, 2, 3, 4\}$. In each of these tasks, we choose a range of ring circumferences that possess optimal (uniform flow) speeds in the range $[5, 10]$ m/s. The corresponding range of acceptable lengths is chosen to be $[250N_{AV}, 360N_{AV}]$ m. At the start of every rollout, a new length is chosen from this range to ensure that any solution learned does not overfit to a specific desired speed. To simulate the effects of lane changes, we follow the process highlighted in~\cite{wu2017multi} for stochastically introducing and removing vehicles while extending this approach to include arbitrary placements of vehicles within a leading gap during cut-ins. The lane-change interval upon which vehicles enter and exit is considered a hyperparameter and is further discussed in later sections. These simulations are reproducible online from: REDACTED FOR REVIEW PURPOSES.


For all RL experiments in this paper, we use the Trust Region Policy Optimization (TRPO)~\cite{schulman2015trust} policy gradient method for learning the control policy, discount factor $\gamma = 0.99$, and step size $0.01$, and a diagonal Gaussian MLP policy with hidden layers $(64, 64, 64)$ and a ReLU non-linearity. The parameters of this policy are shared among all agents in the execution/evaluation procedure and jointly optimized in the training procedure.


\begin{figure*}
    \begin{subfigure}[b]{0.92\textwidth}
        \begin{subfigure}[b]{0.11\textwidth}
            \begin{tikzpicture}
                \node[] () at (0, 1.3) {\footnotesize $44$ Total, $2$ AV};
                \node[] () at (0, 0.9) {\footnotesize Length $= 500$};
                \node[] () at (0, 0) {\footnotesize \textcolor{white}{.}};
            \end{tikzpicture}
        \end{subfigure}
        \begin{subfigure}[b]{0.21\textwidth}
            \includegraphics[width=\textwidth]{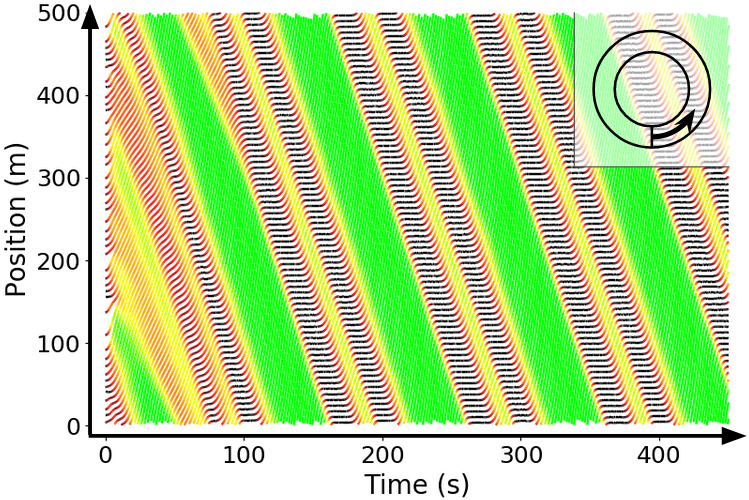}
        \end{subfigure}
        \hfill
        \begin{subfigure}[b]{0.21\textwidth}
            \includegraphics[width=\textwidth]{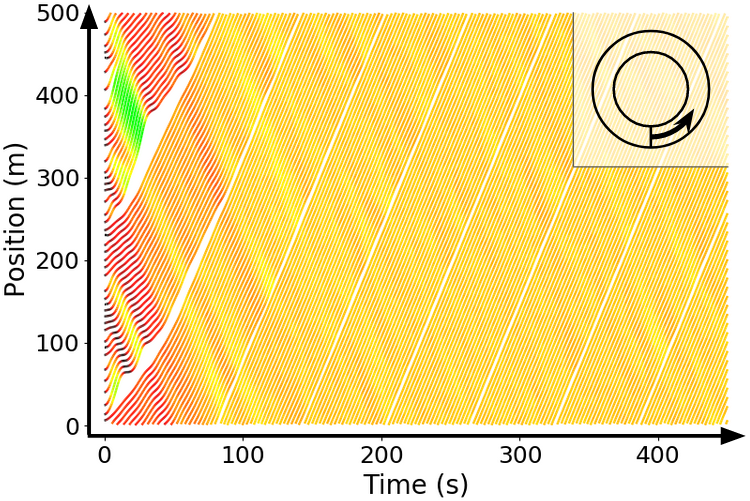}
        \end{subfigure}
        \hfill
        \begin{subfigure}[b]{0.21\textwidth}
            \includegraphics[width=\textwidth]{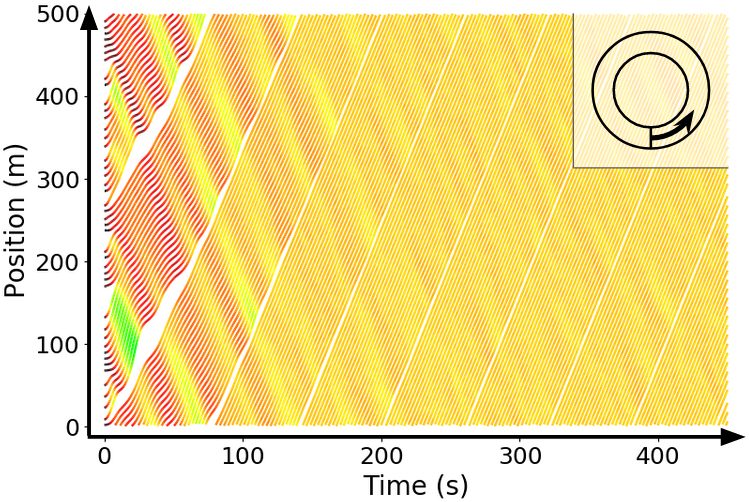}
        \end{subfigure}
        \hfill
        \begin{subfigure}[b]{0.21\textwidth}
            \includegraphics[width=\textwidth]{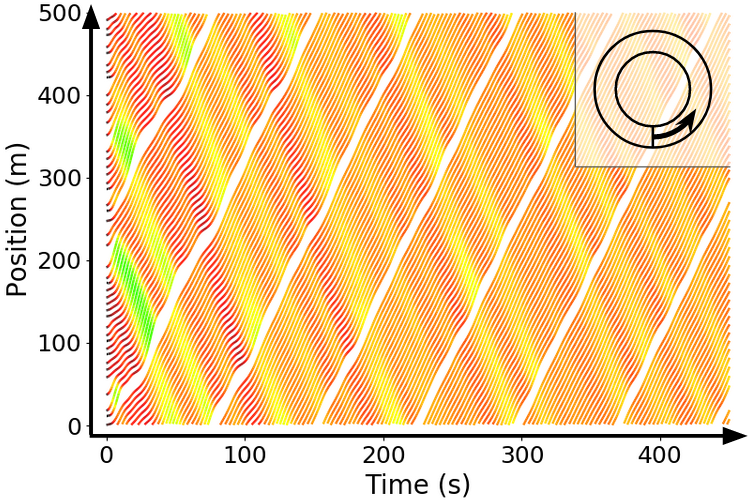}
        \end{subfigure}\\[-5pt]
        \begin{subfigure}[b]{\textwidth}
            \begin{tikzpicture}
                \draw[dashed] (0,0) -- (\textwidth,0);
            \end{tikzpicture}
        \end{subfigure}\\[5pt]
        \begin{subfigure}[b]{0.11\textwidth}
            \begin{tikzpicture}
                \node[] () at (0, 1.3) {\footnotesize $66$ Total, $3$ AV};
                \node[] () at (0, 0.9) {\footnotesize Length $= 750$};
                \node[] () at (0, 0) {\footnotesize \textcolor{white}{.}};
            \end{tikzpicture}
        \end{subfigure}
        \begin{subfigure}[b]{0.21\textwidth}
            \includegraphics[width=\textwidth]{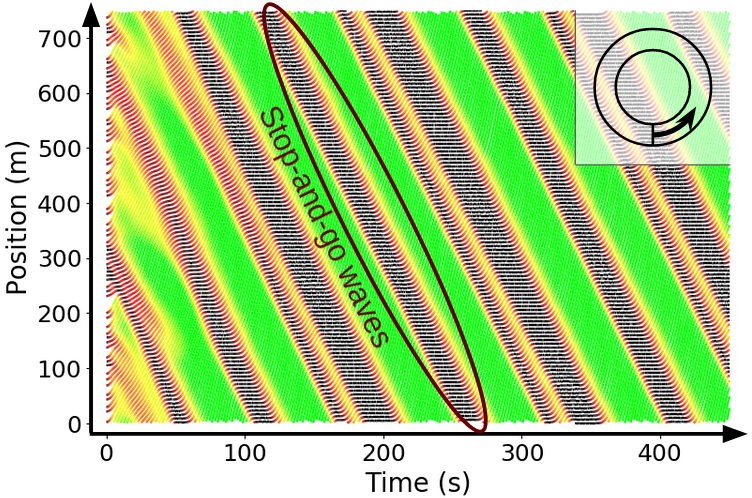}
        \end{subfigure}
        \hfill
        \begin{subfigure}[b]{0.21\textwidth}
            \includegraphics[width=\textwidth]{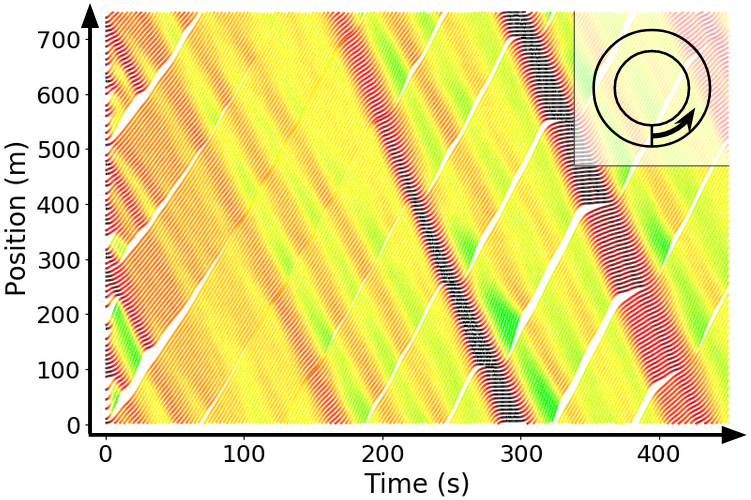}
        \end{subfigure}
        \hfill
        \begin{subfigure}[b]{0.21\textwidth}
            \includegraphics[width=\textwidth]{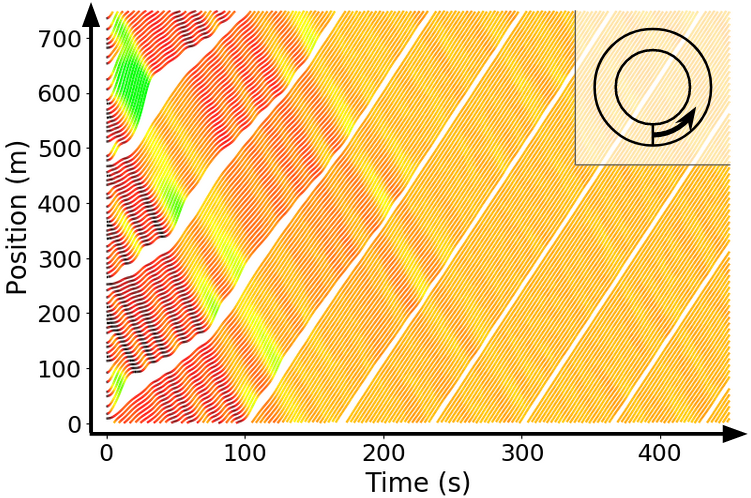}
        \end{subfigure}
        \hfill
        \begin{subfigure}[b]{0.21\textwidth}
            \includegraphics[width=\textwidth]{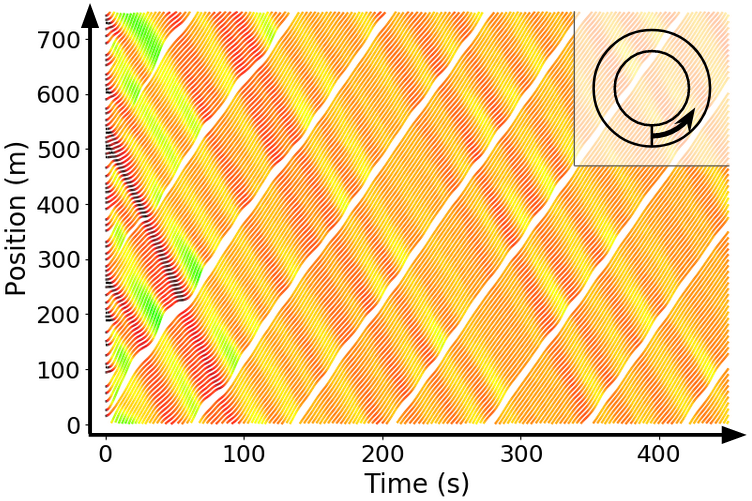}
        \end{subfigure}\\[-5pt]
        \begin{subfigure}[b]{\textwidth}
            \begin{tikzpicture}
                \draw[dashed] (0,0) -- (\textwidth,0);
            \end{tikzpicture}
        \end{subfigure}\\[5pt]
        \begin{subfigure}[b]{0.11\textwidth}
            \begin{tikzpicture}
                \node[] () at (0, 1.8) {\footnotesize $88$ Total, $4$ AV};
                \node[] () at (0, 1.4) {\footnotesize Length $= 1000$};
                \node[] () at (0, 0) {\footnotesize \textcolor{white}{.}};
            \end{tikzpicture}
        \end{subfigure}
        \begin{subfigure}[b]{0.21\textwidth}
            \includegraphics[width=\textwidth]{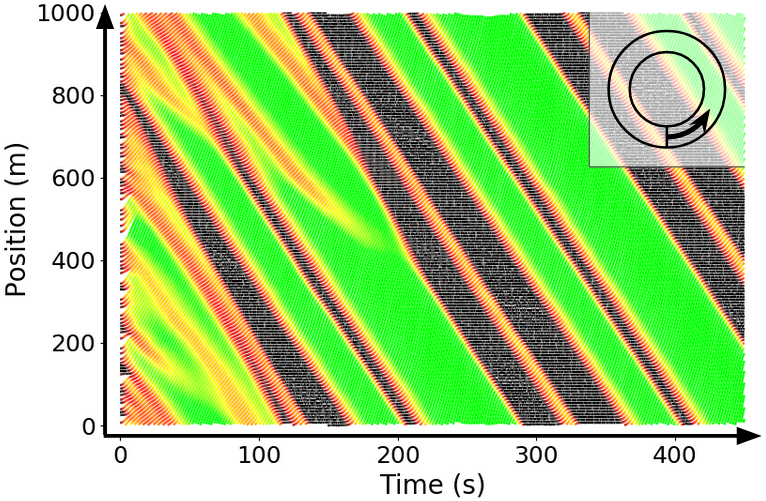}
            \caption{Fully-human baseline}
            \label{fig:ts-baseline}
        \end{subfigure}
        \hfill
        \begin{subfigure}[b]{0.21\textwidth}
            \includegraphics[width=\textwidth]{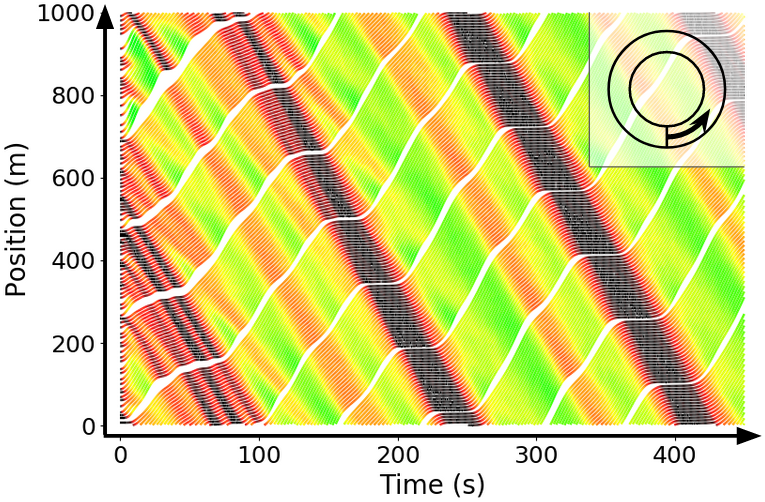}
            \caption{Without curriculum}
            \label{fig:ts-no-curriculum}
        \end{subfigure}
        \hfill
        \begin{subfigure}[b]{0.21\textwidth}
            \includegraphics[width=\textwidth]{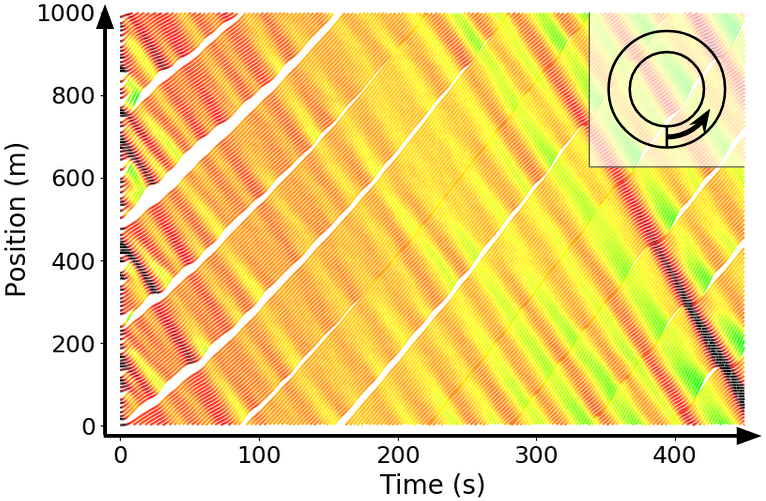}
            \caption{With curriculum}
            \label{fig:ts-curriculum}
        \end{subfigure}
        \hfill
        \begin{subfigure}[b]{0.21\textwidth}
            \includegraphics[width=\textwidth]{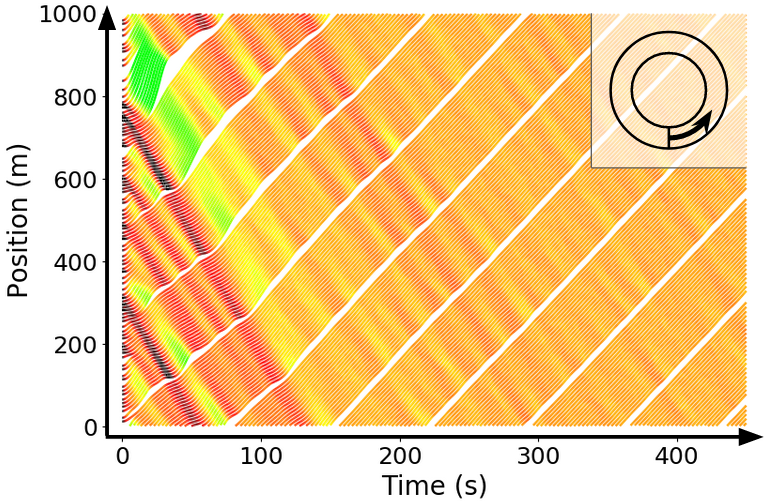}
            \caption{curriculum + stochastic LC}
            \label{fig:ts-curriculum-lc}
        \end{subfigure}
    \end{subfigure}
    \hfill
    \begin{subfigure}[b]{0.06\textwidth}
        \includegraphics[width=\textwidth]{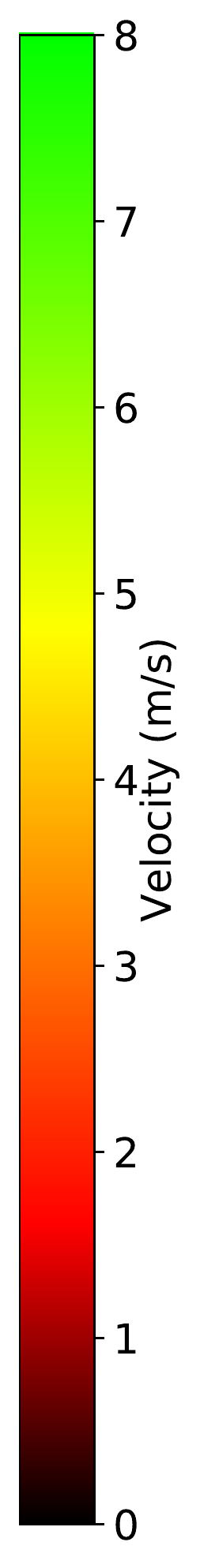}
    \end{subfigure}\\[-7pt]
    \caption{An illustration of the spatio-temporal performance of humans and different learned policies in mixed-autonomy ring roads of different lengths. The use of curricula results in better traffic-smoothing policies in larger ring road settings. This smoothing behaviors performed slightly less efficiently when lane changes are introduced however, due to the increased difficulty of the task.}
    \label{fig:ring-ts}
\end{figure*}
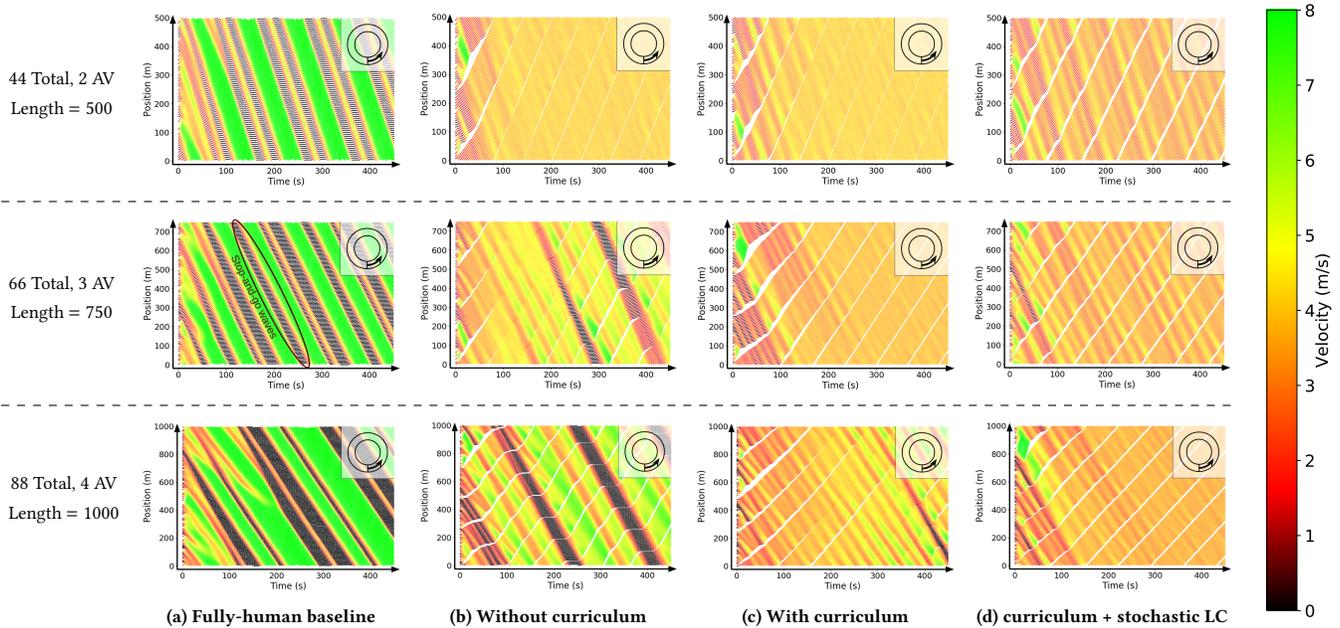

\begin{figure}
    \begin{subfigure}[b]{\linewidth}
        \centering
        \includegraphics[width=\linewidth]{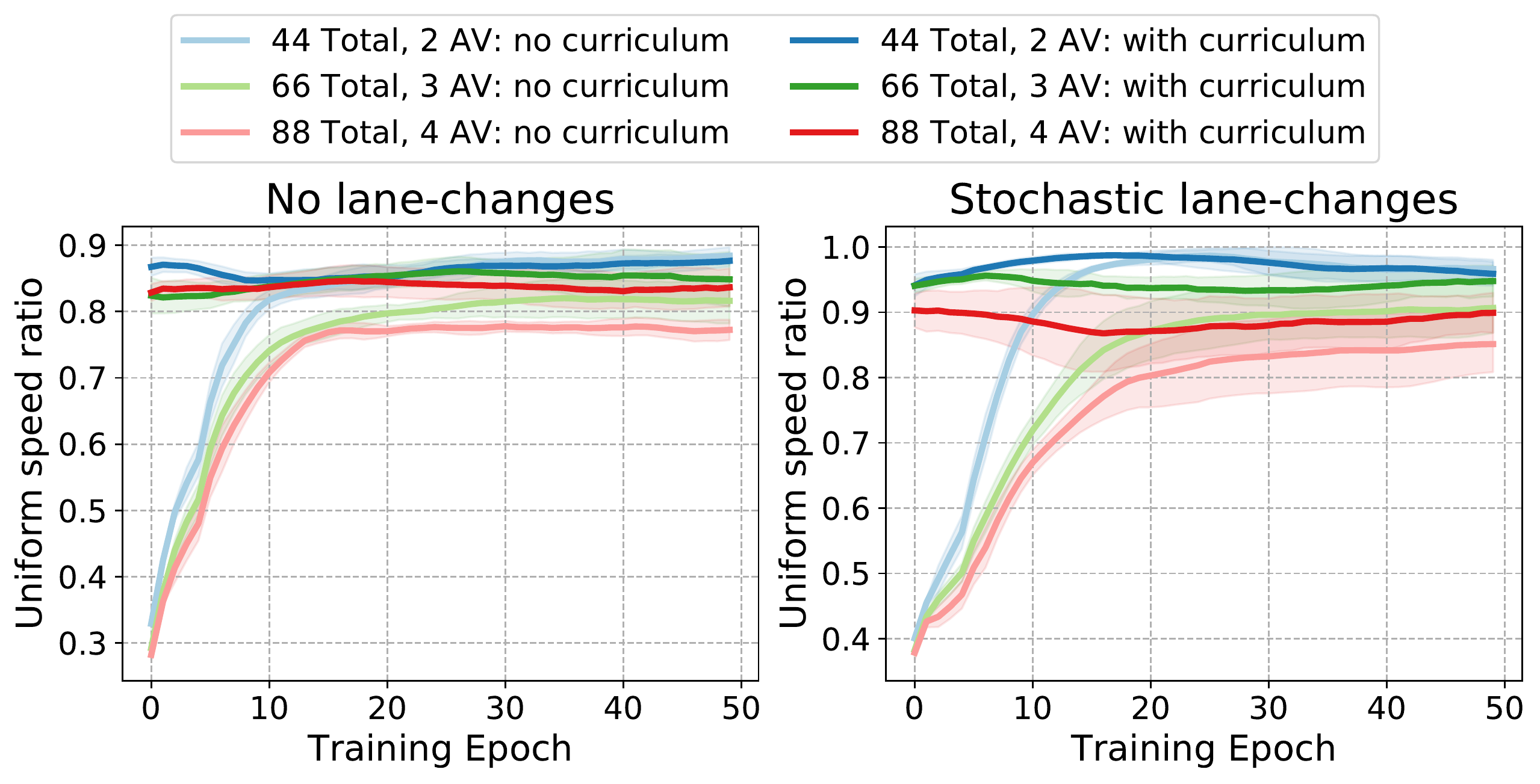}
    \end{subfigure}
    \\[-7pt]
    \caption{Learning performance,
    evaluated on the 
    speed of vehicles in comparison to their uniform flow.
    We find that the use of curricula 
    is
    increasingly beneficial to the learned behavior as the size and number of agents within the problem increases. This is particularly the case when stochastic lane-changes further perturb 
    the system.}
    \label{fig:ring-learning}
\end{figure}

\section{Numerical Results} \label{sec:results}

In this section, we present numerical results for the training procedures presented in the previous section. Through these results, we aim to answer the following questions:

\begin{enumerate}
    \item Does the curriculum learning procedure improve the scalability of decentralized policies learned in larger ring roads?
    \item To what extent does the trained policy succeed in mimicking lane-changing behaviors by human drivers in single-lane settings?
    \item What effect does the utilization of larger rings and dynamics of lane-changing have on the transferability of policies to realistic multi-lane highway networks?
\end{enumerate}

Reinforcement learning experiments for the different setups are executed over $5$ seeds. The training performance is averaged and reported across all these seeds to account for stochasticity between simulations and policy initialization.

\subsection{Ring Experiments} \label{sec:training-performance}

We begin by evaluating the performance of the training procedure and the importance of utilizing curricula in solving larger problems. Figure~\ref{fig:ring-learning} depicts the learning performance on the ring road task with and without the use of stochastic lane changes and curricula. This figure evaluates the policy on its ability to achieve uniform-flow equilibrium over the past $100$ rollout. The evaluation metric $m$ for an individual rollout consists of the ratio between driving speeds $v_i$ by individual vehicles $i$ and the desired uniform flow speed $V_\text{eq}(\cdot)$ as defined in Section~\ref{sec:problem-definition}. Mathematically, this can be written as:
\begin{equation}
    m = \frac{1}{NT}\sum_{t=1}^T\sum_{i=1}^N \frac{v_i}{V_\text{eq}(N/L)}
\end{equation}
where $T$ is the time horizon, $N$ is the number of vehicles, and $L$ is the ring's circumference. As we can see, while the early stage problems perform well without curricula, policies trained on the larger ring problem struggle to achieve similar performance. On the other hand, when trained via a curriculum of gradually growing rings, the learned policies continue to achieve high scores close to the ideal value of $1$ and outperforming their non-curricular counterparts by up to $10$\%. This improvement is particularly evident once lane changes are introduced to the environment, possibly due to the increased stochasticity and complexity that is forced onto the problem.

Figure~\ref{fig:ring-ts} depicts the spatio-temporal performance of the different learned policies in mixed-autonomy ring roads of different lengths. To visualize these behaviors, we use time-space diagrams to visualize the evolution of vehicles across space and time and color their trajectories based on their current speeds to visualize macroscopic mobility and the presence of stop-and-go congestion. In fully human-driven settings (Fig.~\ref{fig:ts-baseline}), instabilities in human driving behaviors result in the formation of stop-and-go waves after a few minutes, depicted by the red diagonal lines in the figures. In larger rings, these waves span more vehicles and propagate more slowly through the network, rendering them more complex problems to solve. Standard RL methods (Fig.~\ref{fig:ts-no-curriculum}) can perform well in training stop-and-go wave dissipating policies for smaller environments but fail in the larger ring settings. On the other hand, the introduction of curricula to the training procedure (Fig.~\ref{fig:ts-curriculum}) helps significantly scale the performance of these algorithms, producing much fewer waves. Finally, when trained in the presence of stochastic lane changes (Fig.~\ref{fig:ts-curriculum-lc}), slightly more waves are experienced due to the increased complexity of the problem.

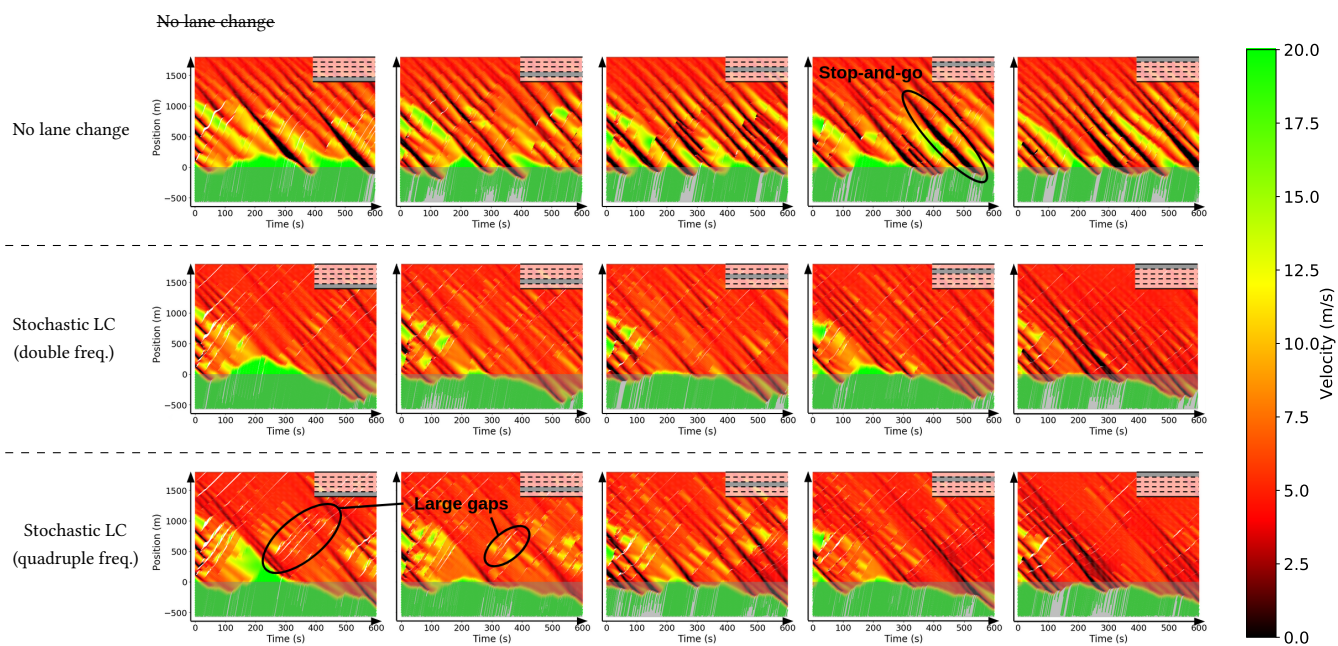
\begin{figure*}
    \begin{subfigure}[b]{0.9\textwidth}
        \begin{subfigure}[b]{0.11\textwidth}
            \begin{tikzpicture}
                \node[] () at (0, 0) {\footnotesize \textcolor{white}{.}};
            \end{tikzpicture}
        \end{subfigure}
        \hfill
        \begin{subfigure}[b]{0.88\textwidth}
            \begin{tikzpicture}
                \node[] () at (0, 0) {\footnotesize \st{No lane change}};
            \end{tikzpicture}
        \end{subfigure}\\[5pt]
        \begin{subfigure}[b]{0.11\textwidth}
            \begin{tikzpicture}
                \node[] () at (0, 1.2) {\footnotesize No lane change};
                \node[] () at (0, 0) {\footnotesize \textcolor{white}{.}};
            \end{tikzpicture}
        \end{subfigure}
        \hfill
        \begin{subfigure}[b]{0.88\textwidth}
            \includegraphics[width=\textwidth]{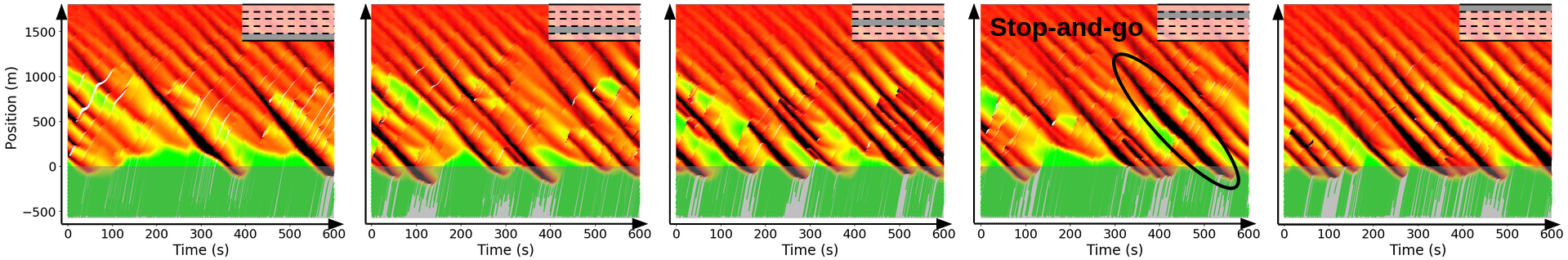}
        \end{subfigure}\\[-5pt]
        \begin{subfigure}[b]{\textwidth}
            \begin{tikzpicture}
                \draw[dashed] (0,0) -- (\textwidth,0);
            \end{tikzpicture}
        \end{subfigure}\\[5pt]
        \begin{subfigure}[b]{0.11\textwidth}
            \begin{tikzpicture}
                \node[] () at (0, 1.4) {\footnotesize Stochastic LC};
                \node[] () at (0, 1.0) {\footnotesize (double freq.)};
                \node[] () at (0, 0) {\footnotesize \textcolor{white}{.}};
            \end{tikzpicture}
        \end{subfigure}
        \hfill
        \begin{subfigure}[b]{0.88\textwidth}
            \includegraphics[width=\textwidth]{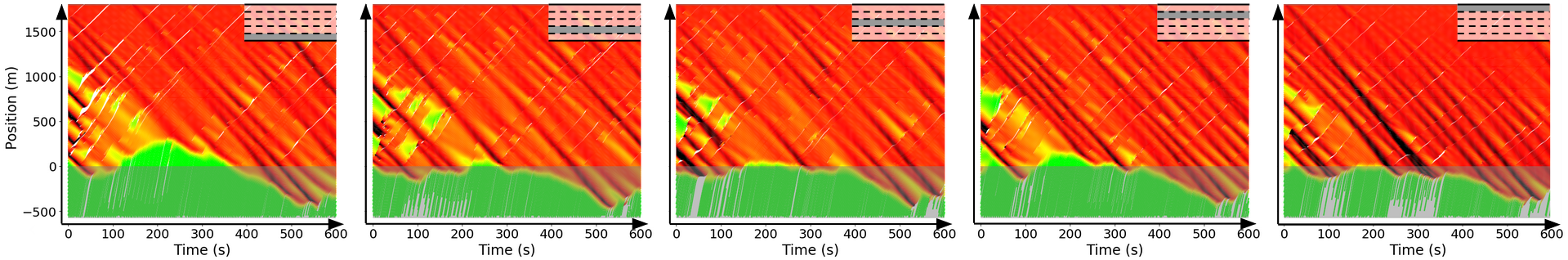}
        \end{subfigure}\\[-5pt]
        \begin{subfigure}[b]{\textwidth}
            \begin{tikzpicture}
                \draw[dashed] (0,0) -- (\textwidth,0);
            \end{tikzpicture}
        \end{subfigure}\\[5pt]
        \begin{subfigure}[b]{0.11\textwidth}
            \begin{tikzpicture}
                \node[] () at (0, 1.4) {\footnotesize Stochastic LC};
                \node[] () at (0, 1.0) {\footnotesize (quadruple freq.)};
                \node[] () at (0, 0) {\footnotesize \textcolor{white}{.}};
            \end{tikzpicture}
        \end{subfigure}
        \hfill
        \begin{subfigure}[b]{0.88\textwidth}
            \includegraphics[width=\textwidth]{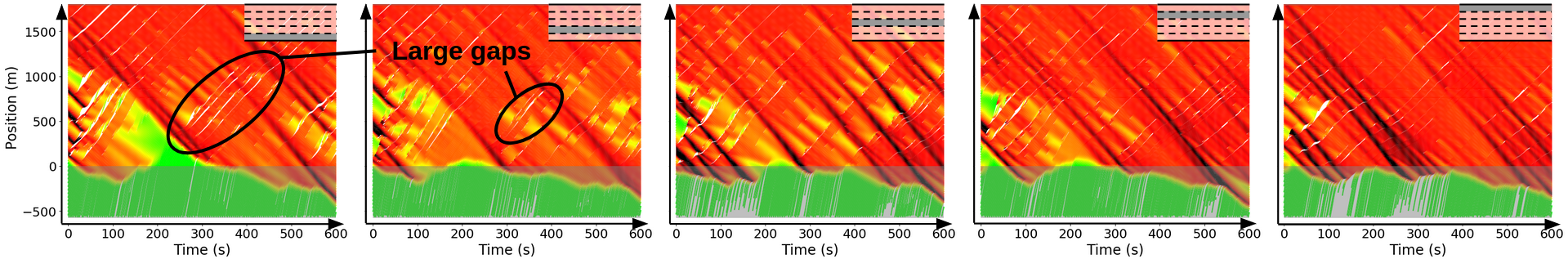}
        \end{subfigure}\\[-7pt]
    \end{subfigure}
    \hfill
    \begin{subfigure}[b]{0.078\textwidth}
        \includegraphics[width=\textwidth]{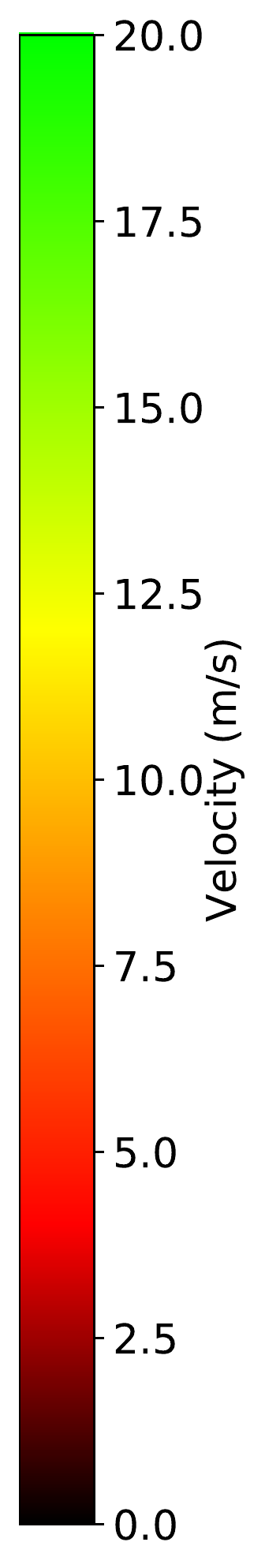}
    \end{subfigure}\\[-7pt]
    \caption{Spatio-temporal performance of the ring policy when transferred to the I-210. The negative positions represent positions outside the controllable regime of traffic where automated vehicles are not ``active'', and instead act as human drivers, so as not to influence the inflow conditions. Policies trained without lane change behaviors do not transfer well and as such produce frequent stop-and-go oscillations. The additional of stochastic lane changes, however, does improve the smoothness of resultant driving, but also results in the formation of unnecessarily large gaps if the lane changes are too frequent.}
    \label{fig:ts-i210}

\end{figure*}

\begin{figure}
    \includegraphics[width=\linewidth]{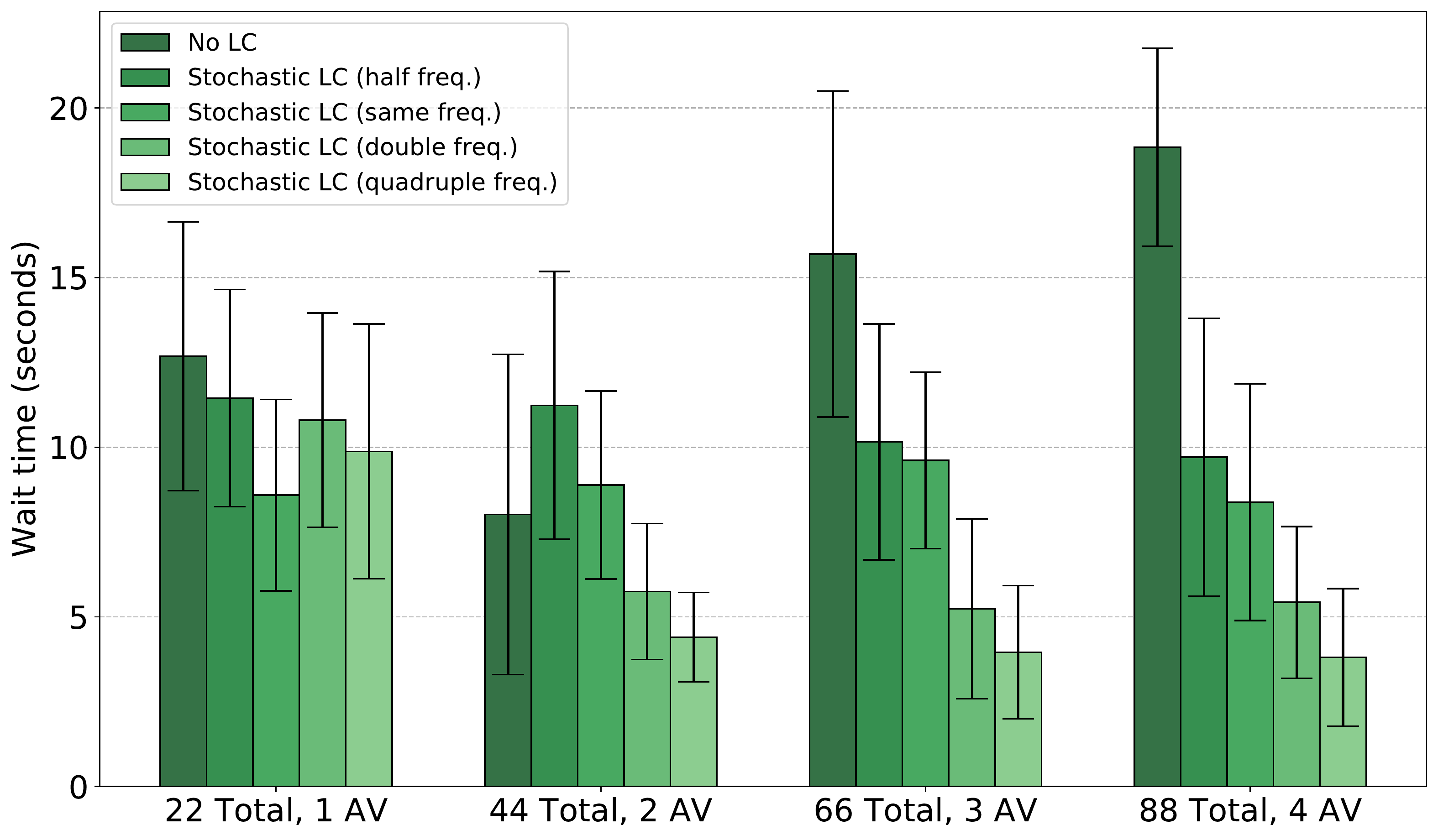}\\[-7pt]
    \caption{Delays incurred by different learned policies when transferred to the I-210. We find that introducing lane changes at increasing frequencies in the source problem reduces in the duration of stopped traffic vehicles experienced in the target network. This is also the case as the size of the problem grows, up to the point at which the boundary no longer play a significant role.}
    \label{fig:i210-delay}
\end{figure}

\subsection{Realistic Network Experiments} \label{sec:transfer-performance}

Next, we validate the generalizability of the learned policy by evaluating its zero-shot transfer to an intricate and calibrated model of human-driving. The target network considered (Figure~\ref{fig:transfer-setup}, right) is a simulation of a $1$-mile section of the I-210 highway network in Los Angeles, California. This network has been the topic of considerable research, with various studies aiming to identify and reconstruct the source of congestion within it~\cite{gomes2004congested,dion2015connected}. We here exploit the work of~\cite{lee2021integrated}, which explores the role of mixed-autonomy traffic systems in improving the energy-efficiency of vehicles within I-210.

Simulations of the I-210 network are implemented in Flow~\cite{wu2021flow}, an open-source framework designed to enable the integration of machine learning tools with microscopic simulations of traffic via SUMO~\cite{krajzewicz2012recent}. These simulations are executed with step sizes of $0.4$~sec/step and are warmed-started for $3600$~sec to allow for the onset of congestion before being run within the evaluation procedure an additional $600$~sec.
Once the initial warmup period is finished, $5$\% of vehicles are replaced with AVs whose actions are sampled from the learned policy to mimic a penetration rate of $5$\%.

Figure~\ref{fig:i210-delay} depicts the performance of the different learned policies once transferred onto the I-210 network. In this task, we can no longer compute the performance as before, as no optimal uniform flow speed exists. Instead, we choose to compute the average time spent stopped by individual vehicles as they traverse the network. This serves as a proxy for the average length and frequency of stop-and-go waves that emerge in the network. Moreover, in choosing an estimation for the expectations in Eq.~\ref{eq:penter}~and~\ref{eq:pexit}, we compute the average number of lane changes experienced in the I-210 in the absence of autonomy and use different magnitudes of this expectation (half, equal/same, double, and quadruple) as the parameter for different algorithms. As we can see from this figure, both the choice of network size and the presence and magnitude of lane changes significantly impact the amount of stopped delay incurred by drivers. From the perspective of magnitude, the transition from $1$ to $2$ AVs captures a significant portion of the improvement, with consecutive gains being marginal. This suggests that the vehicles themselves, when interacting with one another, also largely nullify the effects of the boundary in training time. From the perspective of the lane change frequency, a similar effect is seen as frequencies increase, with the policies learning more cautious maneuvers to avoid the increasing perturbations.

Figure~\ref{fig:ts-i210} depicts the spatio-temporal performance of the learned policy in the $4$-AV setting once transferred to the I-210 network. As noted in the previous paragraph, in the absence of lane changes, policies learned significantly underperforms when transferred to this network. In particular, the policy does not account for the eventuality of waves forming downstream by maintaining a safe distance from its leader, thereby periodically contributing to the propagation of stop-and-go waves. Conversely, the policies trained in the presence of lane changes generalizes nearly perfectly to the new settings, effectively dissipating the vast majority of waves that appear. Similar behaviors seem to occur when the number of lane changes experienced in training is as high as quadruple those in the target task. However, in this latter setting, large gaps begin to form by the automated vehicles in largely uncongested settings. These overly cautious behaviors serve to unnecessarily degrade the mobility of the network, reducing the average speeds of vehicles and the throughput of the entire system despite, interestingly enough, reducing the amount of stopped time experienced on average. This finding suggests that, while exact estimations of lane-change frequencies may not be needed, significant overestimation should be avoided when designing a robust system that is not overly apprehensive of potential perturbations.

\section{Conclusion} \label{sec:conclusion}

This paper explores methods for learning generalizable policies in computationally efficient ring road simulations that effectively transfer to calibrated representations of traffic. First, it introduces a curriculum learning paradigm that utilizes the extendability of the mixed autonomy ring problem to learn scalable interactions between arbitrary numbers of vehicles. Next, it presents a simple method for introducing lane-change style perturbations to the environment in training time. Finally, it demonstrates that combining these two approaches results in a robust learning procedure that produces traffic smoothing policies that generalize to open multi-lane highway networks.

\bibliographystyle{ACM-Reference-Format} 
\bibliography{sample}


\end{document}